\documentclass[aps,12pt, a4paper,nofootinbib]{revtex4}

\usepackage[brazil, english]{babel}
\usepackage[utf8]{inputenc}
\usepackage[T1]{fontenc}
\usepackage{amsmath}
\usepackage{amsfonts}
\usepackage{amssymb}
\usepackage{graphics,graphicx}
\usepackage{ulem}
\usepackage{graphicx,color}
\usepackage{subfigure} %colocar figura lado a lado, figura a) e b).
\usepackage{wrapfig} % pacote reponsavel para colocar figura ao lado do texto
\usepackage{epstopdf}
\graphicspath{{figuras/}}

\usepackage{setspace}
\usepackage[unicode=true,bookmarks=false,breaklinks=false,pdfborder={0 0 1},colorlinks=true]
 {hyperref}
\hypersetup{
 citecolor=blue,linkcolor=blue,urlcolor=blue}
\begin{document}
 
 \title{Thick branes in Horndeski gravity}
\author{Fabiano F. Santos $^{1,2,3}$ and F. A. Brito$^{2,4}$}
\email{fabiano.ffs23@gmail.com, fabrito@df.ufcg.edu.br}
\affiliation{$^1${Instituto de F\'{\i}sica, Universidade Federal do Rio de Janeiro, 21.941-972, Rio de Janeiro - RJ, Brazil.}\\
$^2${Departamento de F\'\i sica, Universidade Federal da Para\'iba, Caixa Postal 5008, 58051-970 Jo\~ao Pessoa PB, Brazil.}\\
$^3${Departamento de Física, Universidade Federal do Maranhão, Campus Universitario do Bacanga, São Luís (MA), 65080-805, Brazil}\\
$^4${Departamento de F\'\i sica, Universidade Federal de Campina Grande, Caixa Postal 10071, 58109-970, Campina Grande PB, Brazil.} }
\date{\today}

\begin{abstract}
We investigate thick brane solutions in the Horndeski gravity. In this setup, we found analytical solutions, applying the first-order formalism to two scalar fields where the first field comes from the non-minimal scalar-tensor coupling and the second is due to the matter contribution sector. With these analytical solutions, we evaluate the symmetric thick brane solutions in Horndeski gravity with four-dimensional geometry. In such a setup, we evaluate the gravity fluctuations to find ``almost massless modes'', for any values of the Horndeski parameters. These modes were used to compute the corrections to the Newtonian potential and evaluate the limit four-dimensional gravity. 
\end{abstract}

\maketitle
\newpage

%\tableofcontents
\newpage

\section{Introduction}

After the seminal works of the Randall-Sundrum \cite{Randall:1999vf,Randall:1999ee} some models were developed to study the gravity localization mechanism in the braneworld \cite{Dvali:2000hr,Karch:2000ct,Bazeia:2004yw,Csaki:2000fc,DeWolfe:1999cp}. Beyond these models, other scenarios were proposed to study solutions through first-order formalism in Einstein gravity \cite{Kiritsis:2013gia,Kiritsis:2019wyk}. However, in recent years, models beyond Einstein's gravity, such as scalar-tensor theories known as Horndeski gravity \cite{DosSantos:2022exb,Santos:2021guj,Brito:2018pwe,Santos:2022lxj,Harko:2016xip,Fu:2019xtx} has been called attention, where such theory have parameters that control the following coupling: $\eta G_{MN}\nabla^{M}\phi\nabla^{N}\phi$, where for sufficiently large $\eta$, the four-dimensional gravity is safely localized on the brane \cite{Brito:2018pwe}. In fact, explicit braneworld solutions can be easily found. For these cases, we restrict ourselves to the analysis of the Newtonian potential. On the other hand, the computation of the weakly excited modes provides the correct potential with an extra term that goes with $1/r^5$, but it is easily suppressed at large distances in comparison with the $1/r$ term, but strongly dominant at small distances. Another interesting advantage of the explicit braneworlds solution for Horndeski gravity is the normalizable bound zero mass gravitational state where wave function shapes the form of the brane in the extra space. For this theory, we have that the continuum of massive modes can produce only very small (negligible) corrections to the Newtonian gravitational potential. This is expected since the analog quantum mechanical potential, uncontrollably grows up at large. The excited modes are separated from the ground state by a gap that is controlled by the Horndeski parameters \cite{Brito:2018pwe}.

 In this paper, we investigate the thick brane system with two scalar fields in the Horndeski theory through first-order formalism, especially the effect of the non-minimal derivative coupling on thick brane \cite{Fu:2019xtx}. First, the equations of motion are presented and a set of analytic background solutions are obtained for the two scalar fields. Then, we investigate the stability of the background scalar profile and with the novel canonically normalized method we show that the original background scalar field is unstable, whereas the canonical one is stable.  In our prescription of two scalar fields in Horndeski gravity, the stability of the thick brane under tensor perturbation is also considered and we show that the graviton zero mode can be localized on the brane. The Newtonian potential on the brane with  $1/r$ scaling is subsequently modified by a repulsion term.

The motivation to study extra dimensions has the aim to probe physics beyond the Standard Model. The recent detection of gravitational waves by collaboration LIGO \cite{Dyson:1920cwa,Abbott:2016blz} and electromagnetic signals from binary systems of compact objects such as neutron stars can help us to constrain the geometry of extra dimensions beyond our known universe with a ($3+1$)-spacetime. For example, in braneworld models, as discussed in \cite{Visinelli:2017bny}, the gravitational waves may propagate through the bulk, taking a `shortcut' through the extra-dimension, while the electromagnetic signals travel on a null-geodesic confined on the brane. As a consequence, we may have a simultaneous event such as the emission of gravitational waves and electromagnetic signals radiation from the merging of binary systems of compact objects that can be detected on the Earth with a time delay due to the different paths of the two signals that propagate with the same speed. Such event in Horndeski theory satisfies $c_{grav}=c_{light}$ in a homogeneous Universe \cite{Creminelli:2017sry,Ezquiaga:2017ekz,Baker:2017hug} and in exact Schwarzschild-de Sitter solution as shown in \cite{Babichev:2013cya,Babichev:2017lmw}.

The paper is organized as follows. In Sec.~\ref{s1} we present the Horndeski gravity. In Sec.~\ref{s2} we develop the first-order formalism in five dimensions and present explicit solutions. In Sec.~\ref{s3} we address the issue of the graviton fluctuations. Finally, in Sec.~\ref{s5} we present our final comments.

% and In Sec.~\ref{s4} we present the holographic scenario through the first order formalism and we analyze the ultraviolet (UV) fixed and the infrared (IR) fixed point by mean the $\beta(\phi)$ function

\section{The Horndeski gravity with a matter scalar field}\label{s1}

In our present investigation, we shall address the study of braneworlds solutions in the framework of the Horndeski gravity which action with a scalar potential reads 
\begin{equation}
I[g_{\mu\nu},\phi]=\int{\sqrt{-g}d^{5}x\left[\kappa(R-2\Lambda)-\frac{1}{2}(\alpha g_{\mu\nu}-\eta G_{\mu\nu})\nabla^{\mu}\phi\nabla^{\nu}\phi-\frac{1}{2}\nabla_{\mu}\chi\nabla^{\mu}\chi-V(\phi,\chi)\right]}.\label{1}
\end{equation}
Note that we have a nonminimal scalar-tensor coupling where we can define a new field ${\phi'}\equiv\Psi$ and $\kappa=1/16\pi G$. In the action (\ref{1}) $G_{\mu \nu}=R_{\mu\nu} -\frac 12 g_{\mu\nu}R$ is the Einstein tensor. This field has dimension of $(mass)^{2}$ and the parameters $\alpha$ and $\eta$ control the strength of the kinetic couplings, $\alpha$ is dimensionless and $\eta$ has dimension of $(mass)^{-2}$. The matter sector is composed of the scalar field $\chi$ that interacts with $\phi$ via the scalar potential $V$. Thus, the Einstein-Horndeski field equations can be formally written in the usual way
\begin{equation}
G_{\mu\nu}+g_{\mu\nu}\Lambda=\frac{1}{2\kappa}T_{\mu\nu}\label{2},
\end{equation}
where $T_{\mu\nu}=T^{(1)}_{\mu\nu}-g_{\mu\nu}V(\phi,\chi)+T^{(2)}_{\mu\nu}$. %The aforementioned energy-momentum tensors $T^{(1)}_{\mu\nu}$ and $T^{(2)}_{\mu\nu}$ take the following form and 
The scalar field equations are given by 
\begin{eqnarray}
\nabla_{\mu}[(\alpha g^{\mu\nu}-\eta G^{\mu\nu})\nabla_{\nu}\phi]=V_{\phi}, \label{3}
\\
\nabla_{\mu}\nabla^{\mu}\chi=V_{\chi}. \label{4}
\end{eqnarray}
We shall adopt the notation $f_{\phi_1\phi_2...\phi_n}(\phi)\equiv d^n f(\phi)/d\phi^1d\phi^2...d\phi^n$. In particular, $V_\phi\equiv{dV}/{d\phi}, V_\chi\equiv{dV}/{d\chi}$.
The aforementioned energy-momentum tensors $T^{(1)}_{\mu\nu}$ and $T^{(2)}_{\mu\nu}$ take the following form
\begin{eqnarray}
&&T^{(1)}_{\mu\nu}=\alpha\left(\nabla_{\mu}\phi\nabla_{\nu}\phi-\frac{1}{2}g_{\mu\nu}\nabla_{\lambda}\phi\nabla^{\lambda}\phi\right)+\nabla_{\mu}\chi\nabla_{\nu}\chi-\frac{1}{2}g_{\mu\nu}\nabla_{\lambda}\chi\nabla^{\lambda}\chi,\label{5}\\
&&T^{(2)}_{\mu\nu}=\eta\left(\frac{1}{2}\nabla_{\mu}\phi\nabla_{\nu}\phi R-2\nabla_{\lambda}\phi\nabla_{(\mu}\phi R^{\lambda}_{\nu)}-\nabla^{\lambda}\phi\nabla^{\rho}\phi R_{\mu\lambda\nu\rho}\right)\nonumber\\
&&-\eta\left((\nabla_{\mu}\nabla^{\lambda}\phi)(\nabla_{\nu}\nabla_{\lambda}\phi)+(\nabla_{\mu}\nabla_{\nu}\phi)\Box\phi+\frac{1}{2}G_{\mu\nu}(\nabla\phi)^{2}\right)\nonumber\\
&&-\eta g_{\mu\nu}\left[-\frac{1}{2}(\nabla^{\lambda}\nabla^{\rho}\phi)(\nabla_{\lambda}\nabla_{\rho}\phi)+\frac{1}{2}(\Box\phi)^{2}-(\nabla_{\lambda}\phi\nabla_{\rho}\phi)R^{\lambda\rho}\right].\label{6}
\end{eqnarray}

% where the Latin indices $M,N=$ $0$, $1$, $2$, $3$ and $5$ run on the bulk and the Greek indices $\mu,\nu=$ $0$, $1$, $2$ and $3$ run along the braneworld.

\section{Equations of motion}
\label{s2}
In this section, we analyze the equations of motion by considering the metric {\it Ansatz} in five dimensions as follows
\begin{eqnarray}
ds^{2}=e^{2A(r)}g_{\mu\nu}dx^{\mu}dx^{\nu}-dr^{2},\label{7}
\end{eqnarray}
For the metric (\ref{7}) we have the following Einstein equation components involving the scalar potential. First, from equation (\ref{2}) the $rr$-component is given by
\begin{equation}
2V(\phi,\chi)+4\kappa(\Lambda+6A^{'2})-\Psi^{2}(r)[\alpha-18\eta A^{'2}]-\chi^{'2}(r)=0.\label{9}
\end{equation}
Another equation can be found from the Einstein equations by combining the $tt$-component with $xx$, $yy$, or $zz$-components to find
\begin{eqnarray}
&&2V(\phi,\chi)+\chi^{'2}(r)+6\eta\Psi(r)\Psi^{'}(r)A^{'}(r)+4\kappa[\Lambda+6A^{'2}(r)+3A^{''}(r)]\nonumber \\
&+&\Psi^{2}(r)[\alpha+6\eta A^{'2}(r)+3\eta A^{''}(r)]=0.\label{10}
\end{eqnarray}
The equations describing the scalar field dynamics come from the equations (\ref{3})-(\ref{4})
\begin{eqnarray}
&&V_{\phi}(\phi,\chi)+4A^{'}\phi^{'}(r)[-\alpha+6\eta A^{'2}(r)+3\eta A^{''}(r)]-(\alpha-6\eta A^{'2}(r))\phi^{''}=0\label{11}\\
&&-V_{\chi}(\phi,\chi)+4A^{'}\chi^{'}(r)+\chi^{''}(r)=0.\label{12}
\end{eqnarray}

\section{First-order formalism}
Let us now reduce the equations of motion to first-order equations by using a superpotential $W(\phi,\chi)$, i.e.,
\begin{eqnarray}
\phi'=\frac12W_{\phi}, \nonumber
\\
\chi'=\frac12W_{\chi},\label{12-first}
\\
A'=-\frac13 W.\nonumber
\end{eqnarray}
Firstly, combining the equation (\ref{9}) with (\ref{10}), for simplicity, we find
\begin{equation}
WW_{\phi\phi}+\frac{W^{2}_{\phi}}{2}+\frac{4}{3}W^{2}-\beta-2\frac{W^{2}_{\chi}}{\eta W^{2}_{\phi}}=0\label{13}
\end{equation}
and now combining the equations (\ref{9}) with (\ref{11}), we have
\begin{equation}
WW_{\phi\phi}+\frac{W^{2}_{\phi}}{2}+\frac{4}{3}W^{2}-\beta-\frac{3}{4\eta}\frac{W_{\chi}W_{\chi\phi}}{WW_{\phi}}=0,\label{14}
\end{equation}
where $\beta=(2\alpha-8\kappa)/\eta$. However, we can see that for consistency when we compare the equations (\ref{13})-(\ref{14}), we get the following constraint on the superpotential 
\begin{equation}
\frac{3}{8}\frac{W_{\chi\phi}}{W}=\frac{W_{\chi}}{W_{\phi}}.\label{15}
\end{equation}
Now choosing the simplest superpotential $W(\phi,\chi)=e^{\sqrt{\frac{8}{3}}(\phi+\chi)}$, we have that the solutions satisfying (\ref{12-first}) consistent with equation (\ref{15}) are given by  
\begin{eqnarray}
&&\phi(r)=-\frac{3}{4}\ln(r),\label{16}\\
&&\chi(r)=-\frac{3}{4}\ln(r),\label{17}\\
&&A(r)=\frac{1}{4}\ln(r).\label{18}
\end{eqnarray}
Besides this superpotential another superpotential that satisfies the equations (\ref{13})-(\ref{14}) is $W=\sqrt{3(\beta+2/\eta)}/4$. This gives the solution $\phi,\chi=const.$ and $A=-(1/3)W_0\, r$, which is the AdS$_5$ vacuum solution. In this regime, Horndeski gravity coincides with Einstein gravity. Furthermore, this solution is well known to localize gravity on the brane sourced by a delta function as long as we patch together the AdS branches $A=\pm kr$ along with the brane according to the solution $A=-k|r|$, where in our case $k=\sqrt{3(\beta+2/\eta)}/4$ is related to the brane tension in the thin wall limit. This is precisely the Randall-Sundrum scenario \cite{Randall:1999vf}. In order to address the issue of the energy distribution of the brane, in a more transparent way, it is interesting to smooth out this brane solution as $k|r|\to \ln{\cosh{kr}}$ \cite{Csaki:2000fc}, such that we have a smooth solution
\begin{equation}
A(r)=-\ln{\cosh{kr}}\approx-\frac{k^2}{2}r^{2},
\end{equation} 
where $k\ll1$ as $\eta\gg1$ (the smooth limit). In our case, we consider the most simple case of the superpotential to satisfy the equations of motion but in future works, we will present other solutions in Horndeski gravity beyond  $W(\phi,\chi)=e^{\sqrt{\frac{8}{3}}(\phi+\chi)}$.

\section{Equation for fluctuations}
\label{s3}
In this section, we study the gravity localization on the brane by considering the solutions obtained with the previously mentioned non-constant constrained superpotential. For this, we focus on tensor perturbations by considering the fluctuations in the metric
\begin{eqnarray}
ds^{2}=e^{2A(r)}(\eta_{\mu\nu}+\epsilon h_{\mu\nu})dx^{\mu}dx^{\nu}-dr^{2},\label{fl.1}
\end{eqnarray}
%where $g_{\mu\nu}=g_{\mu\nu}(\vec{x},r)$ represents the four-dimensional dS, AdS or Minkowski metric, and 
where $h_{\mu\nu}=h_{\mu\nu}(\vec{x},r)$ represents the metric fluctuations with $\epsilon$ a small parameter \cite{Bazeia:2003cv,Gasperini:2007zz,Quiros:2012bh,Llatas:2001jj} . We will perform the computations by considering the first-order perturbations $\delta^{(1)}g_{\mu\nu}=h_{\mu\nu}$, where $h_{\mu\nu}$ is the transverse and traceless (TT) tensor perturbation, i.e., $\partial^{\mu}h_{\mu\nu}=0$ and $h\equiv\eta^{\mu\nu}h_{\mu\nu}=0$ \cite{Karch:2000ct,Csaki:2000fc,DeWolfe:1999cp,Bazeia:2004yw}. Furthermore, since the second scalar field enters into the theory as a matter sector, the tensor fluctuations persist with the usual form \cite{Tsu}. Taking these considerations and the transformation $dr=e^{A}dz$, we have \cite{Santos:2021guj,Fu:2019xtx,Brito:2018pwe}
\begin{eqnarray}
&&C(z)\partial^{2}_{z}h_{\mu\nu}+D(z)\partial_{z}h_{\mu\nu}+\Box_{4d}h_{\mu\nu}=0,\label{T1}\\
&&C(z)=\frac{1-\eta e^{-2A}\phi^{'2}}{1+\eta e^{-2A}\phi^{'2}},\label{T2}\\
&&D(z)=\frac{3A^{'}-\eta e^{-2A}A^{'}\phi^{'2}-2\eta e^{-2A}\phi^{'2}\phi^{''}}{1+\eta e^{-2A}\phi^{'2}}.\label{T3}
\end{eqnarray}
Now, considering the following coordinate transformation $dz=\sqrt{C}d\omega$, we can write for the equation (\ref{T1}):
\begin{eqnarray}
\partial^{2}_{\omega}h_{\mu\nu}+\left(\frac{D}{\sqrt{C}}-\frac{\partial_{\omega}C}{2C}\right)\partial_{\omega}h_{\mu\nu}+\Box_{4d}h_{\mu\nu}=0.\label{T4}
\end{eqnarray} 
By using the decomposition $h_{\mu\nu}(x,\omega)=\epsilon_{\mu\nu}(x)e^{-ipx}H(\omega)$ with $p^{2}=-m^{2}$, we have
\begin{eqnarray}
&&\partial^{2}_{\omega}H(\omega)+Q(\omega)\partial_{\omega}H(\omega)+m^{2}H(\omega)=0,\label{T5}\\
&&Q(\omega)=\frac{D}{\sqrt{C}}-\frac{\partial_{\omega}C}{2C}\label{T6}.
\end{eqnarray}
However, we can simplify the equation (\ref{T5}), by redefining $H(\omega)=G(\omega)\psi(\omega)$. For $G(\omega)=\exp\left(-\frac{1}{2}\int{Q(\omega)d\omega}\right)$, we can compute the Schr\"odinger-like equation as
\begin{eqnarray}
&&-\partial^{2}_{\omega}\psi(\omega)+U(\omega)\psi(\omega)=m^{2}\psi(\omega),\label{T7}\\
&&U(\omega)=\frac{Q^{2}}{4}+\frac{\partial_{\omega}Q}{2}.\label{T8}
\end{eqnarray}
This is an unusual potential as compared with those in the literature \cite{DeWolfe:1999cp,Csaki:2000fc,Bazeia:2004yw,Karch:2000ct}. However, one can easily recover the usual case as $\alpha=1$ and $\eta=0$, otherwise, we have the potential derived from Horndeski gravity. The equation (\ref{T7}) can be factorized as
\begin{eqnarray}
\left(\partial_{\omega}+\frac{Q}{2}\right)\left(-\partial_{\omega}+\frac{Q}{2}\right)\psi(\omega)=m^{2}\psi(\omega),\label{T9}
\end{eqnarray}
where we can explicitly see that there is no tachyon state, since the Schr\"odinger-like operator is factorized in a quadratic positive semi-definite operator. Furthermore, we can find the zero modes by solving the equation (\ref{T9}) setting $m=0$, to find 
\begin{eqnarray}
\psi_{0}=C_{0}\exp\left(\frac{1}{2}\int{Qd\omega}\right)=C_{0}\exp\left(\frac{1}{2}\int{\frac{Qdz}{\sqrt{C}}}\right),\label{T10}
\end{eqnarray}
where $C_{0}$ is a normalization constant. Thus, we have that the normalization condition associated with the graviton zero mode is given by
\begin{eqnarray}
\int{\psi^{2}_{0}(\omega)d\omega}<\infty.\label{T11}
\end{eqnarray}
The problem can be now solved for the solutions \eqref{16}-\eqref{18} % the equation (\ref{T7}) let us first recall the above 
by considering the above transformation of variables between $r$ and $z$ that means $z(r)=\int{e^{-A(r)}}dr$, such that we have
\begin{equation}
A(z)=b+\frac{1}{3}\ln(z);\quad b\equiv\frac{1}{3}\ln\left(\frac{3}{4}\right),\label{T12}
\end{equation}
and also %Through this equation, we have
\begin{eqnarray}
&&C(z)=\frac{z^{8/3}-\eta e^{-2b}}{z^{8/3}+\eta e^{-2b}},\label{T13}\\
&&D(z)=\frac{1}{z}\left(\frac{z^{8/3}-\eta e^{-2b}}{z^{8/3}+\eta e^{-2b}}\right)-\frac{2\eta e^{-2b}}{z^{2}(z^{8/3}+\eta e^{-2b})}.\label{T14}
\end{eqnarray}
However, we can note an interesting aspect of the coefficients (\ref{T13})-(\ref{T14}) in which the potential (\ref{T8}) falls off just slowly enough, i.e., $U(z)\sim z^{-2}$, to rise a normalizable bound-state, that includes the AdS scenario. Thus, for $z\to\infty$ the potential goes to zero and provides a continuum of scattering states $\psi_{m}(z)$ with eigenvalues $m^{2}\geq 0$. In addition, since $dz=\sqrt{C}d\omega$, thus at this limit, we find $C(z)=1$ and $D(z)=1/z$ and as consequence $\omega=z$. Now, we have $U(\omega)=-1/(4\omega^{2})$, and the solutions for equation (\ref{T7}) are given by
\begin{eqnarray}
\psi_{m}(\omega)=A_{m}\sqrt{\omega}J_{0}(m\omega)+B_{m}\sqrt{\omega}Y_{0}(m\omega).\label{T15}
\end{eqnarray}
Following the procedure of \cite{Csaki:2000fc,Bazeia:2003cv}, we can study the localization that occurs for some critical regions of the function argument. For example, for $m\omega>>1$, we have that the Bessel functions (\ref{T15}) become plane waves. On the other hand, the Bessel functions can be power-expanded in $m\omega$ for $m\omega<<1$. Now consider a dimensionful scale $k$ that satisfies $m<<k$, such that several regions, e.g.,  %(but $k\omega>>1$) where $k$ is only a single dimensionful scale. Besides, 
 $\omega>>1/k$ and $m\omega<<1$ can be now studied. After imposing appropriate boundary conditions and normalization function procedure, we can estimate the coefficients $A_{m}$ and $B_{m}$ to obtain the wavefunction at $\omega=0$ given by
%we have $\psi_{0}(\omega)\sim\sqrt{\omega}$ with $A_{m}\sim 1$ and $B_{m}\sim m^{2}$. However, we need to normalize the wavefunction as a plane wave, for this, we must multiply it by an overall factor of $m^{-3/2}$. On the other hand, the value of the wavefunction at $\omega=0$ in order in $m$ is given by:}
\begin{eqnarray}
\psi_{m}(0)\sim\left(\frac{m}{k}\right)^{-3/2}.\label{T16}
\end{eqnarray}
%where the factor $k$ is dictated by dimensional analysis, and $\psi^{2}(0)\sim (m/k)^{-3}$. 
In this case, we have that the corrections to Newton's law are given by:
\begin{eqnarray}
V(r)=G_{N}\frac{M_{1}M_{2}}{r}\left(1+\frac{\bar{C}}{(kr)^{-2}}\right),\label{T17}
\end{eqnarray}
where $\bar{C}$ is a dimensionless number. Since confining gravity has a connection with de Sitter space, one might be tempted to assume that $k$ is somewhat related to the Hubble constant, e.g., $k\sim H_0$. In this scenario, we have that the potential has the correct 4D Newtonian $1/r$ scaling but is subsequently modified by a repulsion term \cite{Dvali:2000hr}.% the repulsion term is associated with scalar extra, and in our case, this term comes from scalar fields $\phi$ and $\chi$ in Horndeski gravity, where this is due the manifest ghost.

\section{Conclusions}
\label{s5}
In this work, we investigated the thick brane system in Horndeski theory where we performed a reduction through the first-order formalism with two scalar fields, especially the effect of the non-minimal derivative coupling on the thick brane model through the $\eta$ parameter that controls the coupling of the scalar field $\phi$. In our studies, the non-minimal derivative coupling, the background scalar field $\chi$ is also non-minimally coupled with the curvature. With these two scalar fields and using first-order formalism, a set of analytic solutions for the brane system was obtained. Such solutions imply that the scalar field $\phi$ is unstable, which means instability for the brane because its superpotential is bottomless. Furthermore, due to the non-minimal coupling for the $\chi$ and the nonminimal derivative coupling for $\phi$, the kinetic term of the scalar field $\phi$ is not canonical and the effective potential of it not only comes from the potential in the action but also from the nonminimal coupling but also due the $\chi$ in the equation (\ref{9}). 

An interesting question in our prescription of the model with two scalar fields is the modification of Newton's law by logarithmic corrections. With this possibility, we need to compensate for additional attraction by using an exchange of a vector particle. But, in our case, we are freer to eliminate this effect that arises due to the second scalar field $\chi$ and can be annihilated considering $\chi=$constant in the action of the theory. Thus, this adjustment for our model is sufficient, however, so that possible equivalence principle violation is not observable. The effect of the scalar field $\chi$ in the theory must be determined by separate phenomenological studies.

\acknowledgments

We would like to thank CNPq and CAPES for partial financial support. FAB acknowledges support from CNPq (Grant no. 309092/2022-1) and PRONEX/CNPq/FAPESQ-PB (Grant no. 165/2018), for partial financial support. This present article arxiv.org/abs/2210.15003 was supported by funded by SCOAP$^3$.

\end{document}